\documentstyle[floats,aps,prl,epsfig,twocolumn]{revtex}

\begin{document}
\bibliographystyle{/usr/share/texmf/tex/latex/revtex/prsty}
\draft
\wideabs{
\title{Modern nuclear force predictions for the $\alpha$-particle}
\author{ 
A.~Nogga\footnote{email:
Andreas.Nogga@ruhr-uni-bochum.de}, H.~Kamada,W.~Gl\"ockle}
\address{Institut f\"ur theoretische Physik II, Ruhr-Universit\"at Bochum,
D-44780 Bochum, Germany }
\date{\today}
\maketitle
\begin{abstract} 
We present new calculations of the $\alpha$-particle which are based on 
the most modern nucleon-nucleon interactions alone and combined with
the Tucson-Melbourne 
or the Urbana IX three-nucleon interaction. Results for the binding
energies and some properties of the wave function are given. On that
phenomenological level little room is left for the action of a
possible four-nucleon force.  
\end{abstract}
}
\narrowtext
Few nucleon bound states have received increasing attention in recent years.
The possibility of  solving increasingly complex systems allows to 
probe the underlying dynamics directly (for a detailed review, see \cite{carlson98}). 
Several methods have been developed and applied 
to the 4N system and realistic forces, 
the GFMC \cite{pudliner97}, 
the CHH \cite{viviani98}, the SV \cite{suzukilec}, the CRCGV \cite{kameyama89} 
and the FY \cite{kamada92,glockle93} method.  Very recently   
a new no core shell model calculation appeared \cite{navratil00}. 
The 4N system  is an important test ground 
for both, the NN and the 3N nuclear interactions, because of its strong binding. 
In this article  we want to 
address the question whether at least the ground state energy of the 
$\alpha$-particle can be  described by the most modern nuclear Hamiltonians.   

In recent years NN forces have been tuned very well to the rich NN data
base, which led to a new generation of so called realistic NN forces:
Nijm~I, II\cite{nijm93}, AV18\cite{av18} and CD-Bonn\cite{cdbonn}. 
While Nijm~II and AV18 are purely
local, Nijm~I has a weak nonlocality in form of a $\nabla^2$-dependence 
and CD-Bonn is quite nonlocal keeping  the underlying
Dirac structure of the 
one-boson-exchange without p/m expansions. All potentials describe
the NN data base with a $\chi^2$/data very close to 1. 
But they also have a
large phenomenological character with a typical number of 40 fit
parameters. The AV18 and CD-Bonn forces
distinguish nn, pp and 
np  interactions and thus include charge symmetry(CSB) and
charge-independence(CIB)  
breaking. The interaction AV18 has in addition a whole set of 
electromagnetic corrections built in. 

 It is well known  that the two 3N bound states, $^3$H and $^3$He, are
theoretically underbound using only NN forces
\cite{friar93,nogga97,wu93,sauer86}. 
We compare our latest theoretical results based on
fully converged Faddeev calculations to the experimental binding energies 
in Table~\ref{tab:1}. The two nonlocal potentials lead to less underbinding
than the local ones. 
  The 3N calculations take the full CSB and CIB  of the NN force
into account including the isospin $T=3/2$ admixtures. In case of
$^3$He and $^4$He the Coulomb interaction is included and, in
addition, we use also the electromagnetic corrections in case of AV18.
Because of an implementation error the results for Nijm~I, II changed
with respect to \cite{nogga97}. Table~\ref{tab:1} includes also the 
strongly model dependent kinetic
energies which are correlated to the NN correlations. The local
potentials have a somewhat harder core \cite{nogga97} leading to a
higher kinetic energy.   
\begin{table*}[tbp]
  \begin{center}
    \begin{tabular}{l|r|r|r|r|r|r} 
Potential&   \multicolumn{2}{c|}{$^3$H} & \multicolumn{2}{c|}{$^3$He}& 
                                                             \multicolumn{2}{c}{$^4$He}  \\
         &$E_B$ [MeV]      & $T$ [MeV] & $E_B$ [MeV] &  $T$ [MeV]  & $E_B$ [MeV]  &  $T$ [MeV] \\
\hline
CD Bonn  & -8.012          & 37.42     & -7.272      &  36.55      &    -26.26    &    77.15   \\
AV18     & -7.623          & 46.73     & -6.924      &  45.68      &    -24.28    &    97.83   \\
Nijm I   & -7.736          & 40.73     & -7.085      &  39.97      &    -24.98    &    84.19   \\
Nijm II  & -7.654          & 47.51     & -7.012      &  46.62      &    -24.56    &    100.31  \\
\hline
Exp.     
         &-8.48            & ---       & -7.72       & ---         &   -28.30     &  ---      
    \end{tabular}
    \caption{3N and 4N binding energies for various NN potentials
together with expectation values $T$ of the kinetic energy.}
    \label{tab:1}
  \end{center}
\vspace{-1cm}
\end{table*}

 There are two additional dynamical ingredients, which
should cure that underbinding, relativistic effects and three-nucleon
forces (3NF). Though one can in principle always find an unitary (but
very complex)  
transformation to a Hamiltonian without any 3NF \cite{polyzou90}, 
we keep  
the given 2N interactions for  practical reasons and because of their
generally accepted physical origin.  
In relation to those forces one or both of the two
mentioned dynamical ingredients are needed. There are still controversies
about the role of relativistic effects \cite{forest99,stadler97b} and
as we think also  open conceptual
questions. Thus we consider here only a strictly nonrelativistic
framework, having however in mind that relativistic corrections should
 finally be added.

The topic of 3NF's is as old as nuclear physics \cite{primakoff39} and based on meson
exchanges various processes have been proposed in the past
(for a review see \cite{robilotta87}). Among them the
Fujita-Miyazawa force \cite{fujita57} with an intermediate $\Delta$ generated by the
exchange of two pions sticks out and is implemented in all modern
3NF models. Here we mention the rather popular $2\pi$-exchange 
Tucson-Melbourne model (TM) \cite{coon79}, the Brazilian version
thereof \cite{robilotta86} and the
Urbana 3NF \cite{pudliner97}. There are also extensions to $\pi$-$\rho$ and $\rho$-$\rho$
exchanges \cite{coon93}. The 2$\pi$-exchange model has been critically reviewed
recently and a modified version, TM', has been proposed in \cite{friar99,huber99}
which satisfies at least chiral symmetry. In this article we use the
TM, TM' and the Urbana IX 3NF's.  

In the TM force enters the strong $\pi$NN vertex function parameterized in the
form of a monopole form factor with a cut-off parameter $\Lambda$. Choosing
it around a generally accepted value one achieves the right order of
the lacking  
binding energy in the 3N system \cite{bomelburg86,wu93,stadler95,friar93}. 
Because there is a strong correlation between the $\alpha$ binding energy,
some 3N scattering observables \cite{witala98c} and the 3N binding energy,  
we fine tune $\Lambda$ to the 3N binding energies of $^3$H
and $^3$He and this separately for each of the four NN forces. 
In this step we have not yet included $T=3/2$
admixtures. We list the fit results in Table~\ref{tab:3}.
Thereby the original TM
parameters for the constants a,b,c and d have been used
\cite{coon81}.
In case of the Nijmegen  interactions we did not adjust $\Lambda$ to
the triton, 
because they do not include a modern specification of the 
$^1$S$_0$ nn force. 
Since  for $^3$H the lack of binding energy is a bit larger than for
$^3$He 
the $\Lambda$ values for $^3$H are slightly larger than for $^3$He.  

Now having 
those 3N Hamiltonians at our disposal we can study the
$\alpha$-particle.
For results referring to older forces see for instance 
\cite{carlson98,kamada92,glockle93}.
We rewrite the Schr\"odinger equation into the Yakubovsky equations
(YE) \cite{yakubovsky67} and thus  
decompose the wave function $\Psi$ into 18 Yakubovsky components
(YC). 
Due to the identity of the nucleons the YC's are not
independent from each other and we can reduce their number to
two: $\psi_1$ and $\psi_2$.  
Then the wave function reads 
\begin{equation} 
\label{wavefu} 
\Psi =  (1 - (1+P) P_{34} ) (1 + P ) \psi_1 + (1+P) (1 + \tilde
P ) \psi_2
\end{equation}
and is expressed with the help of the permutations  
$ P =  P_{12} P_{23} + P_{13} P_{23} $  and  $\tilde P =   P_{13} P_{24}$ 
where $P_{ij}$ are transpositions of particles $i$ and $j$. 
Thus $P$ acts on the
3-body subcluster (123)  and $\tilde P$  interchanges the two two-body
subclusters (12) and (34).
This decomposition is highly advisable for scattering states since  the
boundary conditions can most easily be expressed for YC's. In our case
of a bound state it would
in principle be possible to solve directly the Schr\"odinger equation, 
but the usage of two YC's introduces in a natural manner two kinds of
Jacobi coordinates which accelerates the convergence of a  partial wave
decomposition. 

For 4 identical particles the YE's reduce to two
coupled integral equations 
\begin{eqnarray}
\label{eq:yakueq1}
\psi_{1} & = &  G_0 \ t_{12} \ P \ \left[ 
         (1-P_{34}) \ \psi_{1}+\psi_{2} \right] \\ 
\label{eq:yakueq2}
\psi_{2} & = & G_0 \ t_{12} \ \tilde P \ \left[ 
         (1-P_{34}) \ \psi_{1}+\psi_{2} \right] 
\end{eqnarray}
Here in addition to the permutations the free 4N propagator
$G_0$  
and the NN t-operator $t_{12}$ occur. $t_{12}$ is driven by the NN force
$V_{12}$ through the 
Lippmann-Schwinger equation $t_{12} = V_{12} + V_{12} \ G_0 \ t_{12} $.
In case of 4N scattering one has to go one step further and solve the 
3-body and 2+2 subcluster problems beforehand in order to define the correct
cut-structure \cite{kamada92}. This is not necessary for bound states 
and the form of Eqs. (\ref{eq:yakueq1})-(\ref{eq:yakueq2}) is
easier to handle numerically. One, presumably the most effective
manner, to include 3NF's has been given in \cite{glockle93a}. We
stick to that. 
Then only the first of the two YE's is changed into 
\begin{equation}
\label{yakueq3nf}
\psi_{1} =   G_0 t_{12} P \left[ 
         (1-P_{34}) \psi_{1}+\psi_{2} \right] 
        + (1+G_0 t_{12}) G_0  V_{123}^{(3)} \Psi  
\end{equation}
Here $V_{123}^{(3)}$ is that part of the 3NF which is symmetrical under
exchange of particles 1 and 2. In case of the 2$\pi$-exchange TM or the
Urbana 3NF such a separation into 3 parts is very natural.
We solved the sets Eqs. (\ref{eq:yakueq1})-(\ref{eq:yakueq2})
or Eqs. (\ref{yakueq3nf})-(\ref{eq:yakueq2}) 
in momentum space and in a partial wave
representation. 
The first YC $\psi_1$ stands for the ``3+1'' partition and is
naturally described by two Jacobi momenta for the 3-body subcluster
and one for the relative motion of the 4$^{\rm{th}}$ particle to the other
3. The second YC $\psi_2$ stands for the ``2+2'' partition and is
naturally described by two relative momenta for the inner motion of
the subclusters (12) and (34) and by the relative motion of the two
subclusters. There are a lot of  orbital and spin angular momenta as
well as isospin quantum numbers to be coupled to $J^\pi=0^+$ and
$T=0$. Both basis sets for the ``3+1'' and ``2+2'' partitions comprise
about 1800 different combinations thereof in order to reach a
converged description. The various
momenta are discretized with roughly 35-45 grid points each.  
This leads to a huge absolutely full kernel matrix of dimensions $10^8
\times 10^8$.  
We solve the eigenvalue problem by a Lanczos type algorithm
\cite{saake,stadler91} and make
intensive use of a massively parallel supercomputer.

We would like to remark 
that the introduction of the two kinds of Jacobi momenta in
Eqs. (\ref{eq:yakueq1}),(\ref{eq:yakueq2}) 
and (\ref{yakueq3nf}) leads to additional coordinate transformations
which are hidden in the operator form of the YE's. They are equivalent
to permutations. The crucial point in our calculation is the treatment of
these permutations and coordinate transformations. The direct
interchange of arbitrary particles is unfeasible because 
of the huge dimension of the problem. It is therefore necessary to
interchange particles in two steps in such a way that at least one of
the Jacobi momenta is not changed. This guarantees a block diagonal 
structure for the permutations. Only in this manner the calculation
becomes feasible. For a detailed description see \cite{kamada92}.

Since we allow for CIB and CSB in the NN forces in principle
the dominant total 
isospin state $T=0$ has to be supplemented by $T=1$ and $T=2$ admixtures. Our
estimations lead to the result that their admixtures will change the
binding energy only very slightly ( $<$ 10~keV) 
and thus at this stage we neglected
them. But CIB and CSB lead to the prescriptions that the NN
t-operators occur in the form 
$ t = {1 \over 3} \ t_{np} + {1 \over 3} \ t_{pp}  + {1 \over 3} \ t_{nn} $
in the NN isospin 1 channels.
This is different from the 3N system, where the corresponding linear
combination is
$t = {1 \over 3} \ t_{np} + {2 \over 3} \ t_{pp}  $.
The $pp$ $t$-matrix also includes the effect of the Coulomb
interaction.  Since the bound nucleons are confined to
a small space region, we can put to zero the Coulomb 
interaction outside a radius of $10 - 20$~fm. Then the
Fourier transformation 
of this interaction is nonsingular. The results are cut-off
independent and numerically stable.

\begin{table*}[tbp]
  \begin{center}
    \begin{tabular}{l|r|r|r|r|r|r|r}
Potential&  $\Lambda$ [$m_\pi$] &    
  \multicolumn{2}{c|}{$^3$H} & \multicolumn{2}{c|}{$^3$He}& 
                                                             \multicolumn{2}{c}{$^4$He}  \\
 &       & $E_B$ [MeV]     & $T$ [MeV] & $E_B$ [MeV] &  $T$ [MeV]  & $E_B$ [MeV]  &  $T$ [MeV] \\
\hline
CD Bonn + TM  & 4.784
         & {\bf -8.480}          & {39.10}     & -7.734      &  38.24      &    {-29.15}    &    {83.92}   \\
CD Bonn + TM  & 4.767
         & -8.464          & 39.03     & {\bf -7.720}      &  {38.18}      &    {-29.06}    &    {83.71}   \\
AV18 + TM & 5.156    
         & {\bf -8.476}          & {50.76}     & -7.756      &  49.69      &    {-28.84}    &   {111.84}   \\
AV18 +TM & 5.109    
         & -8.426          & 50.51     & {\bf -7.709}      &  {49.47}      &    {-28.56}    &   {110.92}   \\
AV18 + TM' & 4.756    
         & -8.444          & 50.55     & {\bf -7.728}      &  {49.54}      &    {-28.36}    &   {110.14}   \\
Nijm I + TM & 5.035    
         & -8.392          & 43.35     & {\bf -7.720}      &  {42.59}      &    {-28.60}    &    {93.58}   \\
Nijm II + TM   & 4.975
         & -8.386          & 51.02     & {\bf -7.720}      &  {50.13}      &    {-28.54}    &   {113.09}   \\
\hline
AV18 + Urb IX & ---    
         & -8.478          & 51.28     & -7.760      &  50.23      &    -28.50    &   113.21  \\
AV18+Urb IX (GFMC) \cite{pudliner97}   & ---
         & -8.47(1)          & 50.0(8) & ---         &  ---        &   -28.30(2)  &   112.1(8) \\
AV18+Urb IX (CHH)   \cite{kievskypriv}  & ---
         & -8.476          & 51.26     &             &             &              &            \\
\hline
Exp.     & 
         &-8.48            & ---       & -7.72       & ---         &   -28.30     &  ---      
    \end{tabular}
    \caption{Cut-off parameters $\Lambda$, adjusted 3N binding
energies, resulting $\alpha$ particle binding energies for various force
combinations. Expectation values of the kinetic energy are also
shown. Bold faced results have been adjusted to the experiment.}
    \label{tab:3}
  \end{center}
\vspace{-1cm}
\end{table*}
\begin{table*}[tbp]
  \begin{center}        
    \begin{tabular}{l|r|r|r|r||r|r|r|r}
Model   & \multicolumn{4}{c||} {$^4$He} &  \multicolumn{4}{c} {$^3$He} \\           
& $S$ [\%]  & $S'$ [\%]  &  $P$ [\%]  & $D$ [\%] & $S$ [\%]  & $S'$ [\%]  &  $P$ [\%]  & $D$ [\%]
\\
\hline
AV18                & 85.45     & 0.44       &  0.36      &  13.74 & 89.95     & 1.52       &  0.06      &   8.46     \\
AV18+TM($^3$He)     & 85.10     & 0.30       &  0.75      &  13.84 & 89.86     & 1.26       &  0.15      &   8.72     \\
AV18+TM'($^3$He)    & 83.27     & 0.31       &  0.75      &  15.68 & 89.46     & 1.25       &  0.13      &   9.16    \\
AV18+Urb-IX         & 82.93     & 0.28       &  0.75      &  16.04 & 89.39     & 1.23       &  0.13      &   9.25    \\
CD-Bonn             & 88.54     & 0.50       &  0.23      &  10.73 & 91.45     & 1.53       &  0.05      &   6.98     \\
CD-Bonn+TM($^3$He)  & 89.23     & 0.43       &  0.45      &   9.89 & 91.57     & 1.40       &  0.10      &   6.93      \\
    \end{tabular}
    \caption{$S$, $S'$, $P$ and $D$ state probabilities for $\alpha$
and $^3$He}
    \label{tab:6a}
  \end{center}
\vspace{-1cm}
\end{table*}

\begin{figure}[htbp]
  \begin{center}
    \psfig{file=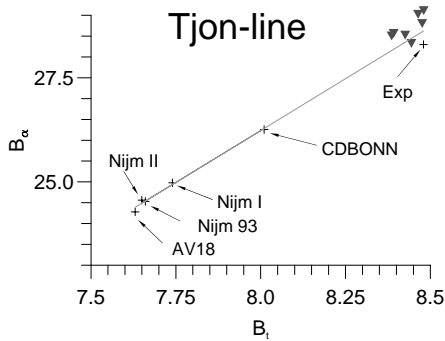,angle=-90,scale=0.25}
    \caption{Correlation of $^4$He against $^3$H binding energies in MeV for
              the different potentials. The triangulars mark the
              predictions with 3NF from Table II.}
    \label{fig:1}
  \end{center}
\vspace{-0.5cm}
\end{figure}

We show in Table~\ref{tab:1} 
the $\alpha$-particle binding energies using 
NN forces only. As expected this theory underbinds the 
$\alpha$-particle. The CD-Bonn result compares well with
\cite{navratil00}.
A correlation between the $^3$H and $^4$He binding energies
found in \cite{tjon75} for simple forces remains valid also for the most 
modern ones. This is depicted in Fig.~\ref{fig:1}. The experimental
value is close to that straight line correlation, which nourishes the
hope that by curing $B_t$ one possibly will also cure $B_\alpha$.  

In Table~\ref{tab:3} we present our results adding the TM 3NF
adjusted individually to the 3N 
binding energies. We find a slight overbinding of about 300~keV for AV18,
Nijm~I and II . In case of CD Bonn the overbinding reaches about
800~keV. 
Note that adjusting the TM 3NF to $^3$H leads to a somewhat
larger overbinding. Nevertheless all these numbers indicate that with
those Hamiltonians one reaches the $\alpha$-particle binding energy rather
closely and there is little room left for
the action of 4N forces. On such a phenomenological level, however,  it is 
not possible to decide about the need of 4N forces. It is always
possible to add another piece of a 3N force which in this case should be
repulsive to reach the $\alpha$-particle binding energy more accurately. 
For instance the $\pi$-$\rho$ exchange 3N force would provide at least 
one more
parameter and both nuclei, $^3$He and $^4$He, could be described. In that
respect the Urbana 3NF is adjusted in such a manner
\cite{carlson98} that in addition to
the 3N binding energy also nuclear matter is taken into account. This
then fixes a repulsive piece in that 3NF, with the result that the
$\alpha$-particle binding comes out essentially right. 
We repeated that calculation
for the Urbana 3NF firstly performed with the GFMC method  
\cite{pudliner97}.
 Our result based on YE's  is given in
Table~\ref{tab:3}  
together
with the previous one. There we also show a triton result by the 
CHH-method \cite{kievskypriv}.
We estimate our numerical accuracy to be about $\pm 3$keV ($\pm
50$keV) for the 3N (4N) system. There appear to be small differences
between our and the GFMC results, especially in the kinetic energies. 

Like in $^3$H one can also separate the $^4$He wave function into $S$,
$S'$, $P$ and $D$-state probabilities. Here $S$ is spatially
symmetric, $S'$ has two-dimensional mixed symmetry and $P$ and $D$ are 
the total $L=1$ and 2 orbital angular momentum parts . We compare the
two nuclei in Tables~\ref{tab:6a}. For $^4$He $S'$ is 
reduced, $P$ somewhat enhanced and probabilities shifted from $S$ to
$D$. It is remarkable that the TM 3NF together with CD-Bonn reduces
the $D$-state probability, in contrast to all other cases. 

In \cite{friar99} it is argued that chiral symmetry requires that the
$c$-term in the TM 3NF should be dropped, leading to a TM' 3NF (one
keeps the remaining constants $b$,$d$ unchanged and $a$ is replace by
$a'=a-2c$). As seen in Table~\ref{tab:3} the resulting 
$\alpha$-article 
binding energy after fitting the cut-off to $^3$He coincides
essentially with the $\alpha$-particle binding energy. It is also
interesting to see that the $c$-term has a significant effect on the
wave function as demonstrated in Table~\ref{tab:7}. There we show the
expectation values of the four different terms contributing to the TM
3NF evaluated with the wave function including TM and the $a$,$b$ and
$d$
expectation values based on the wave function including TM'. In
relation to that latter wave function one can also evaluate the
$c$-term. Interestingly it turns out to be repulsive, whereas for the
TM wave function it is attractive. Also it is interesting to see that
for TM the $a$-term is negligible, whereas it has some importance for
TM'. These examples demonstrate again that 3NF's change wave functions
and cannot be treated perturbatively, a fact known since long time
\cite{bomelburg86}.
\begin{table}[tbp]
  \begin{center}
    \begin{tabular}{l|r|r|r|r}
Model               & $a$-term     & $b$-term  &  $c$-term  & $d$-term
\\
\hline
AV18+TM($^3$He)     &   0.003 & -4.56 &  -1.26 &   -1.72    \\
AV18+TM'($^3$He)    & -0.26   & -3.90 &  +3.00 &   -1.02   \\
    \end{tabular}
    \caption{Expectation values (in MeV) of the four parts of the TM 3NF
      model with respect to wave functions generated with AV18+TM and
      AV18+TM' for $^4$He.}
    \label{tab:7}
  \end{center}
\vspace{-1cm}
\end{table}

Summarizing, after adjusting 3NF's to 3N binding energies the 
$\alpha$-particle binding energies based on modern nuclear forces 
are rather close to the experimental value. This indicates that 4N
forces (at least for $T=0$) will be unimportant. 
More details on wave function properties
will  be published elsewhere. 
The FY equations are perfectly well under control for realistic NN and 
3NF's and will be a perfect tool to study upcoming new force
structures given in chiral perturbation
theory \cite{kolck94,epelbaum00}.  

This work was supported financially by the Deutsche
Forschungsgemeinschaft (A.N. and H.K.). The numerical calculations
were  performed on the CRAY T3E of the NIC in J\"ulich.
\vspace{-0.5cm}

\bibliography{/net/tp2/home4/andreasm/Work/tex/literatur/literatur}

\end{document}